\documentclass[12pt]{article}
\usepackage{geometry}
\usepackage{amsmath}
\usepackage{amssymb}
\usepackage{amsmath}
\usepackage{amsthm}
\usepackage{bbold}
\usepackage{xspace}
\usepackage{graphicx}

\def\appendix#1{
  \addtocounter{section}{1}
  \setcounter{equation}{0}
  \renewcommand{\thesection}{\Alph{section}}
 \section*{Appendix \thesection\protect\indent \parbox[t]{11.715cm} {#1}}
  \addcontentsline{toc}{section}{Appendix \thesection\ \ \ #1}
  }

\renewcommand{\thefootnote}{\fnsymbol{footnote}}

\numberwithin{equation}{section}

\newcommand{\be}{\begin{equation}}
\newcommand{\ee}{\end{equation}}
\newcommand{\ba}{\begin{aligned}}
\newcommand{\ea}{\end{aligned}}

\def\ebb{\delta E^{{\rm Bethe}}}
\def\m1{\left(-1\right)^{F_i}}

\makeatletter
\def\sla@#1#2#3#4#5{{%
  \setbox\z@\hbox{$\m@th#4#5$}%
  \setbox\tw@\hbox{$\m@th#4#1$}%
  \dimen4\wd\ifdim\wd\z@<\wd\tw@\tw@\else\z@\fi
  \dimen@\ht\tw@
  \advance\dimen@-\dp\tw@
  \advance\dimen@-\ht\z@
  \advance\dimen@\dp\z@
  \divide\dimen@\tw@
  \advance\dimen@-#3\ht\tw@
  \advance\dimen@-#3\dp\tw@
  \dimen@ii#2\wd\z@  \raise-\dimen@\hbox to\dimen4{%
    \hss\kern\dimen@ii\box\tw@\kern-\dimen@ii\hss}%
  \llap{\hbox to\dimen4{\hss\box\z@\hss}}}}
\def\slashed#1{%
  \expandafter\ifx\csname sla@\string#1\endcsname\relax
    {\mathpalette{\sla@/00}{#1}}%
  \else
    \csname sla@\string#1\endcsname
  \fi}
\makeatother

\begin{document}


\thispagestyle{empty}
\begin{flushright}\footnotesize
\texttt{hep-th/0610250}\\
\texttt{AEI-2006-081}\\
\texttt{CALT-68-2613}\\
\texttt{ITEP-TH-52/06}\\
\texttt{UUITP-14/06} \vspace{0.8cm}
\end{flushright}

\renewcommand{\thefootnote}{\fnsymbol{footnote}}
\setcounter{footnote}{0}

\begin{center}
{\Large\textbf{\mathversion{bold} How Accurate is the Quantum String Bethe Ansatz? }\par}

\vspace{1.5cm}

\textrm{Sakura Sch\"afer-Nameki$^{\alpha}$, Marija
Zamaklar$^{\beta}$ and Konstantin Zarembo$^\gamma$\footnote{Also at
ITEP, Moscow, Russia}} \vspace{8mm}

\textit{$^{\alpha}$ California Institute of Technology\\
1200 E California Blvd., Pasadena, CA 91125, USA } \\
\texttt{ss299@theory.caltech.edu} \vspace{3mm}

\textit{$^{\beta}$ Max-Planck-Institut f\"ur Gravitationsphysik, AEI\\
Am M\"uhlenberg 1, 14476 Golm, Germany}\\
\texttt{marzam@aei.mpg.de} \vspace{3mm}

\textit{$^{\gamma}$ Department of Theoretical Physics, Uppsala University\\
751 08 Uppsala, Sweden}\\
\texttt{Konstantin.Zarembo@teorfys.uu.se} \vspace{3mm}


\par\vspace{1cm}

\textbf{Abstract}\vspace{5mm}
\end{center}

\noindent
We compare solutions of the quantum string Bethe equations
with explicit one-loop calculations in the sigma-model on
$AdS_5\times S^5$. The Bethe ansatz exactly reproduces the spectrum
of infinitely long strings. When the length is finite, we find that
deviations from the exact answer arise which are exponentially small
in the string length.

\vspace*{\fill}

\setcounter{page}{1}
\renewcommand{\thefootnote}{\arabic{footnote}}
\setcounter{footnote}{0}




\section{Introduction}

The string sigma-model in $AdS_5\times S^5$ \cite{Metsaev:1998it} is
integrable \cite{Bena:2003wd} and is presumably solvable by means of a Bethe
ansatz
\cite{Kazakov:2004qf,AFS,S,Beisert:2005fw}.
Ultimately we want to understand closed strings with periodic
boundary conditions over a finite range of the world-sheet
coordinate. This is not an easy task from the Bethe ansatz
perspective. It is usually simpler to solve the theory on an
infinite line, when one can define asymptotic states and use
bootstrap to find the scattering matrix \cite{Zamolodchikov:1978xm}.
The diagonalization of the S-matrix then determines the spectrum via
the asymptotic Bethe equations. Such Bethe equations are approximate
for a system of finite size. They do not capture effects of vacuum
polarization by particles that travel around the circle \cite{AJK}.
Circumventing this problem is in principle possible, but requires
the use of more complicated algebraic techniques
\cite{Bazhanov:1996aq}.

The currently known Bethe equations for quantum strings in
$AdS_5\times S^5$ \cite{AFS,Beisert:2005fw} are of this asymptotic
type. They are determined by the S-matrix
\cite{Beisert:2005tm} and would have a chance to be
exact only if interactions on the world-sheet were ultra local,
which is not the case: the scattering states are arguably
 solitons of finite size (giant
magnons) \cite{HM} and we also expect that bare point-like
interactions are smeared over a finite range by vacuum polarization.
This Casimir-type effect is expected to produce exponential
corrections to the energy levels in the large-volume limit
\cite{AJK}. Exponential terms were indeed seen in the one-loop
energy shifts for macroscopic spinning strings \cite{SZ,SS} and in
the dispersion relation of a single giant magnon \cite{AFZ}.

Our goal is to understand how good an approximation the
asymptotic Bethe ansatz for strings of finite length is. To do that we
shall study quantum corrections to a specific class of spinning
string solutions described in appendix A. One-loop corrections to
these solution are known \cite{PTT} and were compared to the
predictions of the Bethe ansatz in our previous paper \cite{SZZ}.
The discrepancies found there were finite rather than exponential.
It was later realized that the Bethe equations themselves are
modified by quantum effects \cite{BT}. The one-loop correction
factor
was found in \cite{HL}
and we will update our calculation to take this factor into account.

 In string theory, the range of the world-sheet coordinate is
 a gauge-dependent quantity. It
should not be confused with the proper size of the string in the
target space, which can be very small even for strings with infinite
world-volume. However, in any physical gauge (light-cone, temporal,
or the like) the internal length of the string is naturally
identified with the target-space momentum measured in the units of
$\alpha '$ \cite{Goddard:1973qh}: The string grows in size when the
momentum becomes large \cite{Polchinski:2001ju}. For the states that we
shall consider the length is $2\pi \mathcal{J}=2\pi J/\sqrt{\lambda
}$, where $J$ is the angular momentum on $S^5$ (dual to the R-charge
in $\mathcal{N}=4$ super-Yang-Mills theory) and $\lambda =g_{\rm
YM}^2N$ is the SYM 't~Hooft coupling. In the decompactification
limit $\mathcal{J}\rightarrow \infty $, one is left with the sigma
model on a line with the coupling constant $2\pi /\sqrt{\lambda}$.
$1/\sqrt{\lambda }$ plays the role of the loop counting parameter in
the sigma-model. The usual perturbation theory then yields a power
series in $1/\sqrt{\lambda }$ for the energy spectrum:
$E=\sqrt{\lambda }\,\mathcal{E}+\delta E+\ldots $.

\section{Finite-size corrections}

The energy, as a function of the string length, can be expanded at
$\mathcal{J}\gg 1 $ as \cite{BT,SZ,AJK,SS}\footnote{Here we
concentrate on the one-loop quantum correction to the energy,
leaving aside the classical part.}
\begin{equation}
\label{e-shift} \delta E = \sum_{l=2}^\infty
\frac{f_l}{\mathcal{J}^l} + \sum_{s=0}^\infty a_{s}\,{\rm e}\,^{-
2\pi s\mathcal{J}} \,.
\end{equation}
It is known that the string Bethe ansatz reproduces all orders in
$1/\mathcal{J}$ (all $f_l$'s) exactly
\cite{SZZ,BT}\footnote{Incidentally, the $1/\mathcal{J}$ expansion
resembles perturbative series in $\lambda $ since $1/\mathcal{J}^2=
\lambda /J^2$. However, fractional powers of $\lambda $ also appear
starting from $O(1/\mathcal{J}^5)$ ("2.5 loops") \cite{BT}.}. For
the reasons explained in the introduction we expect exponential
corrections to also arise.

To see how that happens, let us begin with a simple example: the
zero-point energy of $N$ massive bosons and fermions in two
dimensions with twisted boundary conditions
\begin{equation}\label{sumgen}
 \mathcal{F}\left(\mu _i\right)=\frac{1}{2}\sum_{n=-\infty }^{+\infty }
 \sum_{i=1}^{N}\left(-1\right)^{F_i}
 \sqrt{\left(n+\gamma _i\right)^2+\mu _i^2}
 -\frac{1}{2}\sum_{i=1}^{N}\m1\gamma _i^2.
\end{equation}
 The sum converges if
\begin{equation}\label{converg}
 \sum_{i}^{}\m1 =0,\qquad \sum_{i}^{}\m1\mu _i^2=0.
\end{equation}
In addition, the summation should be performed symmetrically in
$n\rightarrow -n$ if any of the $\gamma _i$'s is non-zero. In what
follows we will only encounter the 
periodic boundary conditions, which correspond to all $\gamma _i=0$,
in which case we denote the zero-point energy by $\mathcal{F}_p$.
But it is instructive to consider a more general twisted sum
(\ref{sumgen}).

The parameters $\mu _i$ in (\ref{sumgen}) are masses measured in the
units of length: $\mu _i=m_i L/2\pi $. Hence the large-volume limit
is $\mu _i\rightarrow \infty $.   The summation over mode numbers
can then be replaced by momentum integration to a first
approximation:
\begin{equation}\label{macr}
 \mathcal{F}\approx \frac{1}{2}\int_{-\infty }^{+\infty }
 dp\,\sum_{i}^{}\m1\sqrt{\left(p+\gamma _i\right)^2+\mu _i^2}
 -\frac{1}{2}\sum_{i}^{}\m1\gamma _i^2
 =-\frac{1}{2}\sum_{i}^{}\m1\mu _i^2\ln\mu _i.
\end{equation}
As expected, the macroscopic part of the zero-point energy does not
depend on the boundary conditions.

The 
approximate expression (\ref{macr}) is obviously exact only in the
infinite-volume limit. The finite-size corrections can be taken into
account by Poisson resummation
\begin{equation}\label{}
 \mathcal{F}=\frac{1}{2}\sum_{s=-\infty }^{+\infty }
 \int_{-\infty }^{+\infty }
 dp\,\,{\rm e}\,^{2\pi ips}
 \sum_{i}^{}\m1\sqrt{\left(p+\gamma _i\right)^2+\mu _i^2}
 -\frac{1}{2}\sum_{i}^{}\m1\gamma _i^2,
\end{equation}
which yields
\begin{equation}\label{gener}
 \mathcal{F}=-\frac{1}{2}\sum_{i}^{}\m1\mu _i^2\ln\mu _i
 -\sum_{s=1}^{\infty }\frac{1}{\pi s}
 \sum_{i}^{}\m1\cos(2\pi s\gamma _i)\mu _iK_1(2\pi s\mu _i).
\end{equation}
In the 
case of periodic boundary conditions this becomes
\begin{equation}\label{ramond}
 \mathcal{F}_{p}=-\frac{1}{2}\sum_{i}^{}\m1\mu _i^2\ln\mu _i
 -\sum_{s=1}^{\infty }\frac{1}{\pi s}
 \sum_{i}^{}\m1\mu _iK_1(2\pi s\mu _i).
\end{equation}
 Since the modified Bessel function behaves at large values of the argument as
\begin{equation}\label{bessel}
 K_1(x)\approx \sqrt{\frac{\pi }{2x}}\,\,{\rm e}\,^{-x},
\end{equation}
 the finite-volume corrections are exponential as long as all modes
 are massive.

\section{Semi-classical strings}

The one-loop energy shift for any rigid-string solution is given by
the sum over frequencies of fluctuation modes
\begin{equation}
\label{full} \delta
E=\frac{1}{2}\sum_{n}^{}\left(-1\right)^{F_n}\omega _n.
\end{equation}
We consider a specific class of solutions which are characterized by
angular momentum $J$ and winding number $m$ on $S^5$ and by spin $S$
and winding number $k$ in $AdS_5$. The parameters of the solution
are related by
\begin{equation}\label{}
 kS+mJ=0.
\end{equation}
One of the winding numbers must be negative, and for definiteness we
choose $m>0$ and $k<0$. The explicit form of the solution is
described in Appendix A.

The frequencies of normal modes for this solution were computed in
\cite{PTT} and are rather complicated functions of
$\mathcal{J}=J/\sqrt{\lambda }$ and $\mathcal{S}=S/\sqrt{\lambda }$.
Half of the bosonic frequencies is known only implicitly as the
roots of a particular fourth-order polynomial. In order to proceed
we make a further simplifying assumption. Namely, we consider the
limit\footnote{If one in addition takes $\mathcal{J}\rightarrow
\infty $ such that $\mathcal{J}/k$ is kept finite, $\delta E$
vanishes to the leading order \cite{MTT}. It would be interesting to
investigate finite-size corrections in this limit as well.}
\be\label{LkLimit} k\rightarrow \infty\,,\qquad \quad
\mathcal{S}\rightarrow 0\,,\qquad \quad \mathcal{J},\, m {~-~\rm
finite}\, \ee and systematically drop $O(1/k)$ corrections.
 The
solution considerably simplifies in this limit. The sum over
frequencies reduces to the free-field expression considered in the
previous section\footnote{This equation is a simple rewriting of
(4.1) in \cite{SZZ} for even $k$. The latter was obtained from the
sum over string modes \cite{PTT} in the limit $k\rightarrow \infty
$. As argued in \cite{Gromov:2007aq} the fermionic mode numbers in
\cite{PTT} should be shifted by $k/2$, which eliminates the slight
irregularity of the large-$k$ limit observed in \cite{SZZ}, so that
the expression below should be valid for any $k$ if it is large
enough, independently of whether it is even or odd.}
\begin{eqnarray}\label{exacte}
 E&=&J+\sqrt{\lambda }\,m+
 \frac{4\mathcal{F}_{p}\left(\sqrt{\mathcal{J}^2-m^2},
 \mathcal{J}+m;\sqrt{\mathcal{J}(\mathcal{J}+m)},
 \sqrt{\mathcal{J}(\mathcal{J}+m)}\right)}{\mathcal{J}+m}
 \nonumber \\ &&
 +\sqrt{\mathcal{J}m}-m+\left(\mathcal{J}+m\right)
 \ln\frac{\sqrt{\mathcal{J}+m}}{\sqrt{\mathcal{J}}+\sqrt{m}}\,.
\end{eqnarray}
The first two terms represent the classical, $O(\sqrt{\lambda })$
energy of the spinning string. The rest is the $O(1)$ one-loop
quantum correction $\delta E$.
 The spectrum
consists of eight degenerate fermions with mass
$\sqrt{\mathcal{J}(\mathcal{J}+m)}$, four bosons with mass
$\sqrt{\mathcal{J}^2-m^2}$, and four bosons with mass
$\mathcal{J}+m$, and satisfies the convergence conditions
(\ref{converg}).

Using (\ref{ramond}) and (\ref{bessel}) we find that at large
$\mathcal{J}$ the one-loop correction to the energy indeed has the
form (\ref{e-shift}):
\begin{eqnarray}\label{StringEnergy}
 \delta E^{\rm string}&=&\sqrt{\mathcal{J}m}-m+\mathcal{J}\ln
 \frac{\mathcal{J}^2}{\sqrt{\mathcal{J}+m}\left(\sqrt{\mathcal{J}}+\sqrt{m}\right)
 \left(\mathcal{J}-m\right)}+m\ln\frac{\sqrt{\mathcal{J}}-\sqrt{m}}{\sqrt{\mathcal{J}+m}}
 \nonumber \\ &&-\frac{2\left(1-\,{\rm e}\,^{-\pi m}\right)^2}{\pi
 \sqrt{\mathcal{J}}}\,\,{\rm e}\,^{-2\pi \mathcal{J}}+\ldots.
\end{eqnarray}
The
first line is the infinite-length limit of the string quantum
correction. It is obtained by replacing the sum over the string
modes by the momentum integral. The exponential term can be
extracted from the first Bessel function in  (\ref{ramond}).


\section{Quantum string Bethe ansatz}

The solution we consider belongs to the $sl(2)$ sector,  Bethe
equations for which read \cite{S,Beisert:2005fw}
\begin{equation}\label{Bethe}
 \left(\frac{x_k^+}{x_k^-}\right)^J
 =\prod_{j\neq k}^{}
 \frac{x_k^--x_j^+}{x_k^+-x_j^-}\,\,
 \frac{1-\frac{1}{x_k^+x_j^-}}{1-\frac{1}{x_k^-x_j^+}}
 \,\,{\rm e}\,^{i\theta (x_k,x_j)}.
\end{equation}
The state with spin $S$ is characterized by $S$ Bethe roots $x_k$.
All Bethe roots are real.  $x_k^\pm$ are defined by
\begin{equation}\label{xpm}
 x^\pm+\frac{1}{x^\pm}=x+\frac{1}{x}\pm\frac{2\pi i}{\sqrt{\lambda
 }}\,.
\end{equation}
The Bethe equations contain a dressing phase of the following
general form \cite{AFS}
\begin{equation}\label{}
 \theta (x_k,x_j)=\frac{1}{\pi }\sum_{r,s=\pm}^{}rs\left(\chi (x^r_k,x^s_j)-\chi
 (x^r_j,x^s_k)\right),
\end{equation}
where the function $\chi (x,y)$ is defined as power series
\begin{equation}
\label{chi}
 \chi (x,y)=-\frac{\sqrt{\lambda }}{2}\sum_{r=2}^{\infty }
 \sum_{n=0}^{\infty }
 \frac{c_{r,n}}{\left(r-1\right)\left(r+2n\right)}\,\,
 \frac{1}{x^{r-1}y^{r+2n}}
\end{equation}
The coefficients $c_{r,n}$ are known to the first two orders in
$1/\sqrt{\lambda }$ \cite{AFS,HL,Beisert:2006ib}
\begin{equation}\label{cis}
 c_{r,n}=\delta _{n\,0}-\frac{8}{\sqrt{\lambda }}\,\,
 \frac{\left(r-1\right)\left(r+2n\right)}{\left(2r+2n-1\right)\left(2n+1\right)}\,.
\end{equation}
The one-loop term was proposed in \cite{HL} and passes a number of
consistency tests: it is universal for all sectors
\cite{Freyhult:2006vr}\footnote{The dressing factor originates from
an overall phase of the S-matrix and thus should be the same for all
string states, which was explicitly checked in
\cite{Freyhult:2006vr}.}, and it satisfies \cite{AF} the
crossing-symmetry relation \cite{Janik}.

In the scaling limit of large $J$, $S$, and $\lambda $, the number
of Bethe roots goes to infinity, but each $x_k$ remains finite. The
distance between $x_k$ and $x_{k+1}$, however, goes to zero as
$1/\sqrt{\lambda }$, so that the Bethe roots form a continuous
distribution, which can be characterized by the density
\begin{equation}\label{}
 \rho (x)=\frac{4\pi }{\sqrt{\lambda
 }}\sum_{k=1}^{S}\frac{x_k^2}{x_k^2-1}\,\delta (x-x_k),
\end{equation}
or by the resolvent
\begin{equation}\label{}
 G(z)=\frac{4\pi }{\sqrt{\lambda
 }}\sum_{k=1}^{S}\frac{x_k^2}{x_k^2-1}\,\,\frac{1}{z-x_k}
 =\int_{}^{}dx\,\,\frac{\rho (x)}{z-x} \,.
\end{equation}

In the scaling limit, the Bethe equations reduce to an integral
equation for the density or, for the simplest class of solutions
that we consider here, to an algebraic equation for the resolvent.
These classical Bethe equations can be derived from the equations of
motion of the string, and encode all information about periodic
solutions of the sigma-model \cite{Kazakov:2004qf}. The discreteness
of the quantum Bethe equations leads to an anomalous order
$1/\sqrt{\lambda }$ correction to the classical equations
\cite{Beisert:2005mq}.
The anomaly contribution was computed in our previous work
\cite{SZZ}. Another source of $1/\sqrt{\lambda }$ corrections is the
$O(1)$ term in the dressing phase (\ref{cis}). Taking both
corrections into account
 we
obtain
\begin{equation}\label{algebra}
 G^2(z)-2\pi \left(k-2\,\frac{\mathcal{J}z+m}{z^2-1}\right)G(z) +
 +\frac{4\pi ^2}{z^2-1}\left[k(\mathcal{E}+\mathcal{S}
 -\mathcal{J})z
 -m(k+m)\right]  =
 \frac{1}{\sqrt{\lambda }}\,\,\frac{z^2}{z^2-1}V(z).
\end{equation}
The ``external potential" $V(z)$ is calculated in appendix B
(eq.~(\ref{allbethe})).

Neglecting the right hand side of (\ref{algebra}) we get the
classical solution
\be\label{resolvant}
 G_0 (z)=\pi \left(k-2 {\mathcal{J}z+m \over z^2-1}\right)
 + {\pi \sqrt{P(z)} \over z^2-1}\,,
\ee
where
\begin{equation}
 P(z)=k^2 z^4
 -4k (\mathcal{E}+\mathcal{S}) z^3+2(2\mathcal{J}^2+2m^2-k^2) z^2
 +4k(\mathcal{E}-\mathcal{S}) z
 +k^2 \,.
\end{equation}
Consistency requires that the polynomial $P(z)$ has a double root
\begin{equation}\label{pp'}
 P(c)=0,\qquad P'(c)=0.
\end{equation}
The resolvent then is an analytic function on the complex plane with
a single cut between two other roots of $P(z)$. The density is
defined on this cut and has a typical square-root form: $\rho
(x)\sim\sqrt{(b-x)(x-a)}$. The equations (\ref{pp'}) determine the
energy as a function of charges:
$\mathcal{E}=\mathcal{E}(\mathcal{J},\mathcal{S})$ in a parametric
form. It can be shown that (\ref{pp'}) are equivalent to
(\ref{what'sE}) \cite{SZZ}.

The correction term in (\ref{algebra}) shifts the energy by
$O(\lambda ^0)$. The shift can be calculated from the one-cut
consistency condition on the resolvent. The details of the
calculation can be found in \cite{SZZ}. Here we just give the
result, valid for any external potential $V(z)$:
\begin{equation}\label{}
 \delta E^{\rm Bethe}=\frac{cV(c)}{4\pi ^2k}\,.
\end{equation}
With the explicit form of the potential from (\ref{allbethe}), we
get
\begin{eqnarray}\label{BetheResult}
 \ebb&=&\frac{c}{2\pi ^2k}\int_{}^{}dx\,\,\frac{\rho (x)}{c-x}\,
 \int_{}^{}dy\,\,\frac{\rho (y)}{c-y}\,
 \left\{
 \frac{2}{xy-1}
 +\left[\frac{x-y}{\left(xy-1\right)^2}+
 \frac{1}{x-y}\right]
 \ln\frac{\left(x+1\right)\left(y-1\right)}
 {\left(x-1\right)\left(y+1\right)}
 \right\}
 \nonumber \\
 && -\frac{c}{k}\int_{}^{}dx\,\,
 \frac{\rho '(x)\rho (x)\coth\pi \rho (x)}{c-x}\,.
\end{eqnarray}
This expression is valid even at finite $k$, but is difficult to
handle. Let us now take the large winding limit.

At large $k$ the density is highly peaked at $x\approx -1$, and it
is necessary to introduce the rescaled variable $v$:
\begin{equation}\label{}
 x=-1-\frac{v}{|k|}\,.
\end{equation}
The parameter $c$ behaves as
\begin{equation}\label{}
 c=1+O\left(\frac{1}{k}\right)
\end{equation}
and the density becomes \cite{SZZ}
\begin{equation}\label{}
 \label{RhoLargek} \rho (v) =\frac{|k|}{v
 }\,\sqrt{2\left(\mathcal{J}+m\right)v -v
 ^2-\left(\mathcal{J}-m\right)^2 }\,.
\end{equation}
The single integral in (\ref{BetheResult}) (the second line) can be
done   \cite{SZZ}:
\begin{eqnarray}\label{EBethe}
 \ebb&=& \sqrt{m\mathcal{J}}-\frac{\mathcal{J}+m}{2}\,\ln
 \frac{\sqrt{\mathcal{J}}+\sqrt{m}}{\sqrt{\mathcal{J}}-\sqrt{m}}
 +\frac{1}{4\pi ^2}
 \int_{}^{}\frac{dv}{v}\,\sqrt{2\left(\mathcal{J}+m\right)v -v
 ^2-\left(\mathcal{J}-m\right)^2 }
 \nonumber \\ &&\times
 \int_{}^{}\frac{dw}{w}\,\sqrt{2\left(\mathcal{J}+m\right)w -w
 ^2-\left(\mathcal{J}-m\right)^2 }
 \left[
 \frac{v^2+w^2}{\left(v+w\right)^2\left(v-w\right)}\,\ln\frac{v}{w}
 -\frac{1}{v+w}
 \right] \,.\nonumber \\
&&
\end{eqnarray}
The double integral can be also calculated analytically, leading to
\begin{equation}\label{BBBB}
 \delta E^{\rm Bethe}=\sqrt{\mathcal{J}m}-m+\mathcal{J}\ln
 \frac{\mathcal{J}^2}{\sqrt{\mathcal{J}+m}\left(\sqrt{\mathcal{J}}+\sqrt{m}\right)
 \left(\mathcal{J}-m\right)}+m\ln\frac{\sqrt{\mathcal{J}}-\sqrt{m}}{\sqrt{\mathcal{J}+m}}\,,
\end{equation}
in which we can recognize the non-exponential part of the exact
answer (\ref{StringEnergy}).

\begin{figure}[th]
\centerline{\includegraphics[width=8cm]{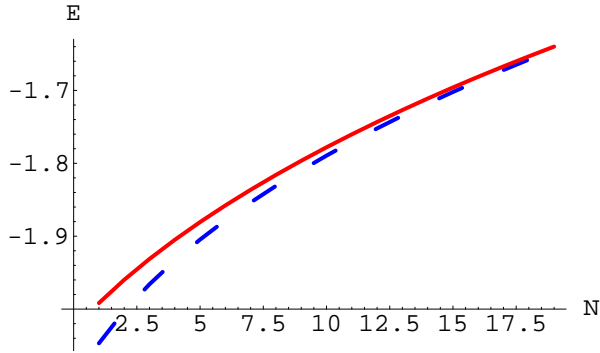}}
\caption{\label{fig1}\small The one-loop correction to the energy as
a function of $\mathcal{J}$ for $m=1$ and ${\mathcal J}= 1 + 0.1 N$.
The blue dashed line is the exact string energy (\ref{exacte}), the
red solid line is the Bethe-ansatz result (\ref{BBBB}). }
\end{figure}

Let us compare (\ref{BBBB}) with the exact energy shift
(\ref{exacte}) numerically. Since the difference is only
exponential, it should be numerically small even for
$\mathcal{J}\sim 1$. But in our case $\mathcal{J}$ cannot be smaller
that $m$. In fact, one of the bosonic modes becomes massless at
$\mathcal{J}= m$ , which means that the correction terms in
(\ref{gener}) cease to be exponentially suppressed. To our surprise,
we found that the difference between (\ref{BBBB}) and (\ref{exacte})
never gets larger that $10\%$ (fig.~\ref{fig1}), even very close to
the massless limit.

\section{Conclusions}

We conclude that the Bethe equations (\ref{Bethe}) are asymptotic
and describe the string spectrum with an exponential accuracy as
long as the string length is sufficiently large. The corrections to
the asymptotic energy levels are of order $\exp(-2\pi
J/\sqrt{\lambda })$. Similar correction arise at weak coupling due
to the wrapping interactions in the SYM \cite{Beisert:2004yq}, which
start to affect the energies (anomalous dimensions) at order
$\lambda ^J=\exp(-\ln(1/\lambda )J)$. It would be interesting to
understand how the exponent in the finite-size corrections
interpolates between $\ln(1/\lambda )$ at $\lambda \rightarrow 0$
and $2\pi /\sqrt{\lambda }$ at $\lambda \rightarrow \infty $.

Can more sophisticated Bethe equations reproduce the exact spectrum
of the closed string with periodic boundary conditions? Literature
on integrable field theories in finite volume is vast:
\cite{Zamolodchikov:1989cf,Bazhanov:1996aq}
contains a necessarily incomplete selection of references. Perhaps
extra, "particle" degrees of freedom, such as those in
\cite{PM,Rej:2005qt}, are necessary to
correctly account for the finite-size effects. Or maybe one should
start from a pseudo-vacuum with all anti-particle levels empty and
then carefully fill the Fermi sea in the finite volume \cite{KZ}. It
is not clear to us what could play the role of the Bethe particles
on the world-sheet, or how to define the pseudo-vacuum in the AdS
string theory, but it would be definitely interesting to repeat the
semiclassical calculations of this paper in  the context of the
truncated models \cite{PM,KZ} for which
these questions have been answered.


\subsection*{Acknowledgements}

We would like to thank R.~Roiban, J.~Teschner and A.~Tseytlin for
interesting discussions, and to N.~Gromov and P.~Vieira for pointing
out subtleties in boundary conditions for fermionic string modes.
The work of K.Z. was supported in part by the Swedish Research
Council under contracts 621-2004-3178 and 621-2003-2742, by the
G\"oran Gustafsson Foundation, and by RFBR grant NSh-1999.2003.2 for
the support of scientific schools. S.S.N. is supported by a Caltech
John A. McCohn Postdoctoral Fellowship in Theoretical Physics.

\setcounter{section}{0}
\appendix{The spinning string solution}

Here we briefly review the string configuration we consider. The
relevant part of the $AdS_5\times S^5$ metric in global coordinates
is
\begin{equation}\label{}
 ds^2=-\cosh^2\rho \,dt^2+d\rho ^2+\sinh^2\rho \,d\theta  ^2
 +d\phi  ^2,
\end{equation}
where the first three terms are the metric of $AdS_3$ and $\phi $ is
the angle of a big circle in $S^5$. The circular string solution
\cite{Arutyunov:2003za} has the following form
\begin{equation}
\begin{aligned}\label{classt}
 \rho  &=\,{\rm const}\,,     & t    &=\kappa \tau ,\\
 \theta&=\sqrt{\kappa ^2+k^2}\,\tau + k\sigma , &
 \phi  &=\sqrt{\kappa ^2+m^2}\,\tau + m\sigma
\end{aligned}
\end{equation}
where
\begin{eqnarray}
 r_1^2\equiv\sinh^2\rho &=&\frac{\mathcal{S}}{\sqrt{\kappa ^2+k^2}}\,,
 \nonumber \\
\mathcal{E}&=&
 \frac{\kappa \mathcal{S}}{\sqrt{\kappa ^2+k^2}}+\kappa \, ,
\end{eqnarray}
and
\begin{eqnarray}\label{what'sE}
 2\kappa \mathcal{E}-\kappa^2&=&2\sqrt{\kappa ^2+k^2}\,\mathcal{S}
 +\mathcal{J}^2+m^2\,, \nonumber \\
 k\mathcal{S}+m\mathcal{J}&=&0.
\end{eqnarray}
The global charges of the string (the energy $E$, the spin $S$, and the
angular momentum $J$) combine with the  string tension into the
following dimensionless ratios, which stay finite in the classical
($\lambda \rightarrow \infty$, $J \rightarrow \infty$, $S\rightarrow
\infty$) limit
$\mathcal{E}=\frac{E}{\sqrt{\lambda }}$,
$\mathcal{S}=\frac{S}{\sqrt{\lambda }}$ and
$\mathcal{J}=\frac{J}{\sqrt{\lambda }}$.

\appendix{Semiclassical Bethe equations}\label{corrections_to_eqs}

In deriving the classical limit of the Bethe equations, the
following integral representation of the dressing phase turns out to
be useful:
\begin{equation}\label{scatt}
 \theta (x_k,x_j)=\frac{1}{\pi }
 \int_{-\frac{2\pi}{\sqrt{\lambda }}}^{\frac{2\pi}{\sqrt{\lambda }}}
 d\varepsilon d\varepsilon'\,\,\frac{x_k^{\varepsilon\,2}
 x_j^{\varepsilon'\,2}  f(x_k^\varepsilon ,x_j^{\varepsilon '})}
 {\left(x_k^{\varepsilon\,2}-1\right)\left(x_j^{\varepsilon'\,2}-1\right)}\,,
\end{equation}
where $x_k^\varepsilon $ is defined by a generalization of
(\ref{xpm}):
\begin{equation}\label{}
 x_k^\varepsilon+\frac{1}{x_k^\varepsilon }=x+\frac{1}{x}+i\varepsilon .
\end{equation}
The function $f(x,y)$ is an anti-symmetrized second derivative of
$\chi (x,y)$ from (\ref{chi}):
\begin{equation}\label{}
 f(x,y)=\frac{\partial ^2\chi (y,x)}{\partial y\,\partial x}
 -\frac{\partial ^2\chi (x,y)}{\partial x\,\partial y}\,.
\end{equation}
Using (\ref{cis}) we get for it
\begin{equation}\label{}
 f(x,y)=\frac{\sqrt{\lambda }}{2}\,\,\frac{x-y}{x^2y^2\left(xy-1\right)}
 +\frac{2}{\left(xy-1\right)\left(x-y\right)}
 +\left[\frac{1}{\left(xy-1\right)^2}+\frac{1}{\left(x-y\right)^2}\right]
 \ln\frac{\left(x+1\right)\left(y-1\right)}{\left(x-1\right)\left(y+1\right)}\,.
\end{equation}

Taking the logarithm of the Bethe equations (\ref{Bethe}), using
(\ref{scatt}) for the dressing phase and similar integral
representations for other terms, we get
\begin{eqnarray}\label{ufh}
 &&\int_{-\frac{2\pi}{\sqrt{\lambda }}}^{\frac{2\pi}{\sqrt{\lambda }}}
 d\varepsilon \,\left\{
 \frac{Jx_k^\varepsilon }{x_k^{\varepsilon \,2}-1}+
 \sum_{j\neq k}^{}\left[
 \left(
 \frac{x_k^{\varepsilon \,2}}{x_k^{\varepsilon \,2}-1}
 +\frac{x_j^{-\varepsilon \,2}}{x_j^{-\varepsilon \,2}-1}
 \right)\frac{1}{x_k^\varepsilon -x_j^{-\varepsilon }}
    \right.\right.\nonumber \\ &&\left.\left.
 -\left(
 \frac{x_k^{\varepsilon}}{x_k^{\varepsilon \,2}-1}
 -\frac{x_j^{-\varepsilon}}{x_j^{-\varepsilon \,2}-1}
 \right)\frac{1}{x_k^\varepsilon x_j^{-\varepsilon }-1}
 \right]\right\}
 \nonumber \\ &&
 -\frac{1}{\pi }\sum_{j\neq k}^{}
 \int_{-\frac{2\pi}{\sqrt{\lambda }}}^{\frac{2\pi}{\sqrt{\lambda }}}
 d\varepsilon d\varepsilon '\,
 \,\frac{x_k^{\varepsilon \,2}x_j^{\varepsilon' \,2}}
 {\left(x_k^{\varepsilon \,2}-1\right)
 \left(x_j^{\varepsilon' \,2}-1\right)}
 \left\{
 \frac{\sqrt{\lambda }}{2}\,\,
 \frac{x_k^\varepsilon -x_j^{\varepsilon'}}
 {x_k^{\varepsilon \,2}x_j^{\varepsilon' \,2}
 \left(x_k^\varepsilon x_j^{\varepsilon'}-1\right)}
  \right.\nonumber \\ && \nonumber \left.
 +\frac{2}{\left(x_k^\varepsilon x_j^{\varepsilon'}-1\right)
 \left(x_k^\varepsilon- x_j^{\varepsilon'}\right)}
 +\left[
 \frac{1}{\left(x_k^\varepsilon x_j^{\varepsilon'}-1\right)^2}
 +\frac{1}{\left(x_k^\varepsilon- x_j^{\varepsilon'}\right)^2}
 \right]
 \ln\frac{\left(x_k^\varepsilon +1\right)\left(x_j^{\varepsilon'}-1\right)}
 {\left(x_k^\varepsilon -1\right)\left(x_j^{\varepsilon'}+1\right)}
 \right\}
 =2\pi k.
\end{eqnarray}
Taking the large-$\lambda $ of this equation is not a trivial
exercise because of the anomaly that arises in the summation over
$j\approx k$. We will not repeat all the steps here. They can be
found in \cite{SZZ}. The only new ingredient compared to \cite{SZZ}
is the one-loop correction to the dressing phase, the $O(1)$ terms
in the double integral. But since this term is non-singular at
$x_k^\varepsilon \rightarrow x_j^{\varepsilon '}$, taking its
strong-coupling limit amounts in just dropping the dependence of the
integrand on $\varepsilon $ and $\varepsilon '$.

Multiplying (\ref{ufh})  by $1/(z-x_k)$, summing over $k$ and
basically repeating the calculation in \cite{SZZ}, we get
\begin{eqnarray}\label{allbethe}
 &&G^2(z)-2\pi \left(k-2\,\frac{\mathcal{J}z+m}{z^2-1}\right)G(z)
 +\frac{4\pi ^2}{z^2-1}\left[kz\left(\mathcal{E}+\mathcal{S}-\mathcal{J}\right)
 -m(k+m)\right] \nonumber \\
 &&=\frac{2}{\sqrt{\lambda }}\,\,\frac{z^2}{z^2-1}\int_{}^{}dx\,\,\frac{\rho (x)}{z-x}
 \int_{}^{}dy\,\,\frac{\rho (y)}{z-y}\left\{
 \frac{2}{xy-1}
 +\left[\frac{x-y}{\left(xy-1\right)^2}+
 \frac{1}{x-y}\right]
 \ln\frac{\left(x+1\right)\left(y-1\right)}
 {\left(x-1\right)\left(y+1\right)}
 \right\}
 \nonumber \\ &&
 -\frac{4\pi ^2}{\sqrt{\lambda }}\,\,\frac{z^2}{z^2-1}
 \int_{}^{}dx\,\,\frac{\rho '(x)\rho (x)\coth\pi \rho (x)}{z-x}\,.
\end{eqnarray}


\bibliographystyle{JHEP}
\renewcommand{\refname}{Bibliography}
\addcontentsline{toc}{section}{Bibliography}

\providecommand{\href}[2]{#2}\begingroup\raggedright\endgroup


\end{document}